\documentclass[twocolumn,showpacs,preprintnumbers,amsmath,amssymb]{revtex4}

\usepackage{graphicx}
\usepackage{dcolumn}
\usepackage{bm}

\begin{document}

\preprint{APS/123-QED}

\title{Chemotactic predator-prey dynamics\/}

\author{Ankush Sengupta, Tobias Kruppa, and Hartmut L\"owen}
\affiliation{%
Institut f\"ur Theoretische Physik II: Weiche Materie, 
Heinrich-Heine-Universit\"at \\
Universit\"atsstrasse 1, D-40225 D\"usseldorf, Germany 
}%


\date{\today}

\begin{abstract}
A discrete chemotactic predator-prey model is proposed in which the prey secrets a diffusing chemical
which is sensed by the predator and vice versa. Two dynamical states
corresponding to catching and escaping are identified and it is shown
that steady hunting is unstable. For the escape process,
the predator-prey distance is
diffusive for short times but exhibits a transient subdiffusive behavior
which scales as a power law $t^{1/3}$ with time $t$ and ultimately
crosses over to diffusion again. This  allows to classify the motility
and dynamics of various predatory bacteria and phagocytes. In particular,
there is a distinct region in the parameter space where they prove to
be infallible predators.
\end{abstract}

\pacs{05.40.-a,87.17.Jj,05.10.Gg}

\maketitle

Phagocytes or predatory microbes are hunting their prey by chemotaxis \cite{chemotaxis1,chemotaxis2}, 
i.e. they sense the concentration of a chemical which is secreted
by the prey and is diffusing through the solution \cite{Poon_review}. Typically the 
predator moves along the steepest gradient of the chemical concentration 
to ultimately find its ejection source. Likewise the prey (for example another microbe) ``smells" 
a secreted chemical from the advancing predator and tries to escape by 
moving along in the opposite direction of its
maximal gradient. This chemotactically coupled  predator-prey
systems are relevant for many biological microorganisms. In fact, there are many
examples of biological relevance for chemotactically coupled predator-prey microorganisms.
To name just a few, common microbial predators and phagocytes are {\it Bdellovibrio\/} \cite{Nunez,Strauch,Dori},
{\it P. aeruginosa\/} \cite{Kato}, {\it D. discoideum} \cite{Clarke,Endres},
lymphocytes \cite{Zigmond} and {\it M. xanthus\/} \cite{Astling,Berleman}.

Previous theoretical investigations have focussed on spatiotemporal pattern formation
in  predator-prey colonies \cite{benJacob,Boraas} which are
typically described by nonlinear reaction-diffusion equations \cite{Keller_Segel,Tsyganov}.
While the latter approaches involve a coarse-grained continuum modelling, there 
are much less  model studies on {\it individual\/} microorganisms. A discrete swarming model
of individual self-propelled particles for bacterial colonies has been proposed by
Csirok et al \cite{benJacob}. This  was elaborated recently by Romanczuk et al \cite{Romanchuk} 
based on a related  individual model of Schweitzer and Schimansky-Geier \cite{Schweitzer}.
Finally individual autochemotactic models have been studied where the microbe follows 
its own diffusing secretion \cite{deGennes_2004,Grima_PRL_2005,Grima_PRE_2006,Sengupta_1}. 
In all of these individual models there is no predator involved, apart
from a recent study \cite{Oshanin} which addressed a lattice model
with no chemicals involved.

Here we propose a discrete model which describes both the predator and the prey individually
and contains explicitly the diffusion of the two chemicals
secreted by the predator and the prey together with the Brownian motion of the latter.
 The deterministic (fluctuation-free) model is analyzed analytically
and by numerical solution which is supplemented by Brownian dynamics computer simulations at finite
temperature. Depending on the model parameters and the initial distance between predator and prey,
 two different dynamical processes are identified which correspond to {\it catching\/} and {\it escaping}, and an unstable 
{\it steady hunting}. By analytical treatment, various scaling laws 
are extracted characterizing and delineating the different regimes. In the absence of noise, 
the mean-square distance between predator and prey scales with different exponents $\alpha$ as a function of time
with a subdiffusive anomalous exponent $\alpha=2/3$ 
for escaping \cite{Golestanian} and a ballistic
behavior $\alpha=0$ for steady hunting.
Brownian motion leads to ultimate diffusion ($\alpha=1$) such that these exponents are transient.
Catching is accompanied by a scaling form of $|t-t_{cap}|^{\alpha}$, with
$\alpha=2/3$, $t_{cap}$ being the capture time.
In principle, our results allow  to map and classify different biological systems 
into the different regimes.

In our discrete predator-prey model, the predator is at position
${\bf r}_1 (t)$ at time $t$, hunting the prey which is at position
${\bf r}_2(t)$ and trying to escape. The concentration field of the chemical
secreted by the predator (prey) at a constant ejection rate $\lambda_{1(2)}$
is denoted by $c_{1(2)}({{\bf r},t})$.
Assuming a gradient sensing scenario for each
microbe in response to the chemical secreted by the other, and taking
into account effective stochastic fluctuations that are associated with
the non-equilibrium self-propulsion mechanism of each, the overdamped
equations of motion for the predator and the prey, respectively reads
\begin{eqnarray}
\gamma_1 {\bf {\dot r}}_1 = +\kappa_1 \nabla c_{2} ({\bf r}_{1},t)
+ {\mbox {\boldmath $\eta$}_{1}}(t)
\label{eom1}\\
\gamma_2 {\bf {\dot r}}_2 = -\kappa_2 \nabla c_{1} ({\bf r}_{2},t)
+ {\mbox {\boldmath $\eta$}_{2}}(t)
\label{eom2}
\end{eqnarray}
Here, $\gamma_{1(2)}$, $\kappa_{1(2)}$ and 
${\mbox {\boldmath ${\eta}$}}_{1(2)} (t)$ are the damping constant
in the medium, the chemical coupling constant (a measure of
the gradient sensing strength) and the effective noise vector
associated with stochastic self-propulsion of the predator (prey),
respectively. The first term on the right, in both equations,
models the systematic contribution of chemotactic response through
simple gradient sensing. We take both $\kappa_{1,2} > 0$, so that
$c_{2}$ acts as a chemoattractant for the predator, while $c_{1}$
is a chemorepellant for the prey.
We model ${\mbox {\boldmath ${\eta}$}}_{i} (t)$
as a Gaussian white noise:
$\langle{\mbox {\boldmath $\eta$}}_{i}(t)\rangle = {\bf 0}$,
$\langle{\mathbf \eta_{i\mu}}(t)
{\mathbf \eta_{j\nu}}(t^{\prime}) \rangle = 
2{\gamma_{i}} \beta^{-1} \delta_{ij}\delta_{\mu \nu}\delta(t-t^{\prime})$.
The Greek indices refer to spatial components, while the Roman
indices are reserved for the microbe's attributes
($i = 1$: predator; $2$: prey). Here, $\beta$ corresponds to an inverse
effective temperature associated with the stochastic fluctuations, such
that $D_{i} = 1/(\gamma_i \beta)$ is the non-chemotactic diffusion constant
of the microbe concerned. Hydrodynamic interaction between the microbes
is neglected.

The diffusion equation of each of the chemicals reads
\begin{equation}
\frac{\partial c_{i}({\bf{r}},t)}{\partial t} + {\bf{u}}_{i}({\bf r},t)
\cdot \nabla c_{i}({\bf{r}},t) = 
D_{ci}{\nabla}^2 c_{i}({\bf{r}},t) 
+ \lambda_{i} \delta[{\bf r}-{\bf r}_{i}({t})] 
\label{eq:rd}
\end{equation}
where $D_{ci}$ is the diffusivity of the corresponding chemical, and
we have assumed each microbe as a point-source emitter.
${\bf{u}}_{i}({\bf r},t)$ is the advective flow-field set in the medium
due to the motion of the microbe. In practice, a microbe has a typical
mesoscopic size `$a$'. Expressing lengths in units of $a$:
$r \to r^{\prime\prime} = r/a$, and time in units of $\tau_0 = a^2 / D_{ci}$:
$t \to t^{\prime\prime} = t/\tau_{0}$, Eq.(3) reduces to
\begin{equation}
\left[\frac{\partial}{\partial t^{\prime\prime}}
+ {\frac{a {\bf{u}}_{i}({\bf r},t)}{D_{ci}}} \cdot {\nabla^{\prime\prime}}
- {\nabla^{\prime\prime}}^2 \right] c_{i}({\bf{r}},t) =
{\frac{a^2 \lambda_{i}}{D_{ci}}} \delta[{\bf r}-{\bf r}_{i}({t})]
\label{eq:rd1}
\end{equation}
Simple spatial gradient sensing microbes are known to move slowly with
velocity on the order
$v \sim 10^{-2}-10^{-1} \mu$m/s, and are typically of size $a \sim 1-10 \mu$m.
Chemicals are secreted typically at
$\lambda_i \sim 10^3$ molecules/s and diffuses at
$D_{ci} \sim 10^2-10^3 \mu$m$^2$/s.
Under such practical situations, noting that the magnitude of the flow-field
$|{\bf u}_i|$ can be at most on the order of $v$,
we have the dimensionless factor
$a v/ D_{ci} \sim 10^{-5}-10^{-2}$. This makes the advective term negligible,
and reverting to the original variables: $r^{\prime\prime} \to r$
and $t^{\prime\prime} \to t$, Eq.(4) simplifies to
\begin{equation}
\frac{\partial c_{i}({\bf{r}},t)}{\partial t} 
- D_{ci}{\nabla}^2 c_{i}({\bf{r}},t) =
\lambda_{i} \delta[{\bf r}-{\bf r}_{i}({t})]
\label{eq:rd2}
\end{equation}
Fast moving microorganisms with $v \sim 10-10^2 \mu m/s$
(capable of producing appreciable advection in the medium)
are known to chemotax by `temporal sensing' mechanism of the chemical
gradient \cite{Berg} which we do not address here.
Also, the detailed influence of the microbe's
shape and distribution of chemical sensors on the spatial gradient sensing
itself are not considered in our model.

For an unconfined
space in three dimensions, the Green's function solution to Eq.(5) yields
\begin{equation}
	c_{i}({\bf r},t) = \lambda_{i}\int_{0}^{t}dt'\frac{1}{(4\pi D_{ci}\left|t-t'\right|)^{\frac{3}{2}}} \ \exp\left(\frac{-({\bf r}-{\bf r}_{i}(t'))^2}{4D_{ci}\left|t-t'\right|}\right) \label{eq:green_sol}
\end{equation}
Brownian dynamics of the predator-prey system is implemented to
simulate the chemotactic motion (eqs.(1) and (2)), and using Eq.(6)
to calculate the spatial gradient of the chemical concentration.
We measured time in units of $\lambda_2^{-1}$, lengths in units of
$l_0 =0.1\sqrt{D_{c2}/\lambda_2}$ and energy in units of
$\epsilon_0 = \gamma_2 D_{c2}$. Real estimate
yields $\kappa_i \sim 10^3 \epsilon_0 l_0^3$
for Dictyostelium \cite{Endres} moving at $0.2 \mu$m/s up cAMP gradient of
$0.01$nM$/\mu$m, secreted at $10^3$molecules/s with diffusivity
$300 \mu$m$^2/$s. Microglial cells \cite{Luca}
moving at $2 \mu$m/min in
response to an interleukin gradient of $0.003$nM/$\mu$m secreted at
$200$ molecules/min and diffusing at $900 \mu$m$^2/$min, has
$\kappa_i \sim 10 \epsilon_0 l_0^3$.


\begin{figure}[]
\includegraphics[width=10.0cm]{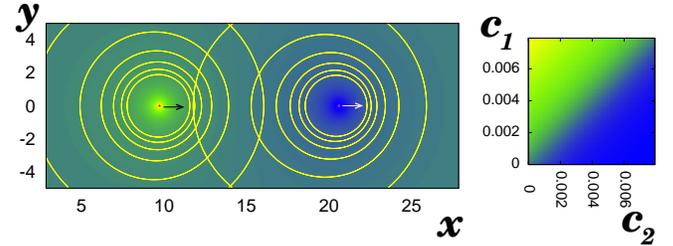}
\caption{\label{figChase} (color online). (Left) Chemotactic Chase: A predator
(red dot on left) chases a prey (red dot on right), while the latter tries to
escape through chemotactic gradient sensing of the diffusing chemicals.
The arrows indicate their respective direction of motion in absence of
fluctuations. The contours around each microbe represent the
equi-concentration lines of the secreted chemicals in a two-dimensional
projected plane in this case, indicating the asymmetry
of the distribution. The color code
used here for the spatial distribution of the secreted chemorepellant ($c_1$)
and the chemoattractant ($c_2$), as they mingle in space, is shown in the
right panel.
}
\end{figure}

A first look at the problem suggests that the ultimate fate of the prey,
i.e.\ whether it will manage to escape, get captured or be steadily hunted
forever, will depend not only on the emission rates and diffusivities
of the chemicals, the effective coupling strengths
of the microbes to the chemical and their mobilities in the medium,
but also on the initial separation between them:
$r_0 = {r}_{12}(t=0) \equiv |{\bf r}_2 (0) - {\bf r}_1 (0)|$.
Having set the model, it is therefore instructive to examine the zero-noise
(${\mbox {\boldmath $\eta$}}_{i}(t) = {\bf 0}$) 
deterministic case first, in order to understand the combination
of variables relevant in predicting the outcome.
Fig.1 shows a simulation snapshot of the chemotactic chase process
in the absence of fluctuations.
Assuming a steady-state velocity $v_i$ for microbe $i$, moving along
the $x$-axis,
the concentration profile for the chemical secreted by it simplifies to
\begin{equation}
c_i ({\bf r},t=0) = ({\lambda_i}/{2 \pi D_{ci}r})\exp\left(-{v_i (x+r)}/{2D_{ci}}\right)
\end{equation}
In the presence of chemotactic coupling,
one then expects $\gamma_i v_i = (1-\delta_{ij})\kappa_i [{\partial c_j (\bf r)} / {\partial x} ]_{{\bf r}={\bf r}_{ij}}$, for each of them. In addition,
demanding a condition for steady hunting, whence $v_i = v_j$,
maintaining a constant separation
$r_{12} = \Delta$, leads to
\begin{equation}
\delta(\Delta^*) = (1+{\Delta^*}^{-1})\exp(-{\Delta^*}^{-1})
\end{equation}
Here, $\Delta^* = \Delta/\Delta_0$,
$\Delta_0 = {\kappa_1 \lambda_2}/({4 \pi D_{c1} D_{c2} \gamma_1})$
being a length scale, and
$ \delta = ({\kappa_1 \gamma_2 \lambda_2 D_{c1}})/({\kappa_2 \gamma_1
\lambda_1 D_{c2}})$
will be termed as a {\em sensibility ratio} in the predator-prey relationship.
Fig.2(a) shows the resulting phase diagram.
For a given $\Delta^*$, if the sensibility ratio is increased,
there is a transition from the free to the trapped phase. To understand
this we note that an increase in the chemo-attractant coupling or
its emission or a decrease of the prey's mobility
would prove advantageous to the predator in sensing the prey at a given
distance. Also, a decrease in the chemoattractant diffusivity will enable
the predator to easily track its prey and ultimately trap it. Similarly,
for a given $\delta$, increase in the separation distance will
be advantageous to the prey in escaping. Further, since
$\Delta > 0$ (predator follows prey), the validity of Eq.(8)
requires $\delta < 1$.  This means the phase boundary between the
trapped and escaped state lies below the $\delta(\Delta^*) = 1$ line in the
parameter space. For $\delta \geq 1$, there is {\em no escape}. Whatever is
the separation between the microbes, the predator will ultimately
capture the prey in this case.

\begin{figure}
\includegraphics[width=10.0cm]{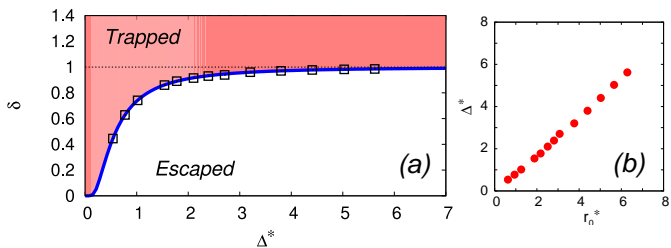}
\caption{\label{figPhaseDiag} (color online).(a) Dynamical Phase Diagram
of the chemotactic predator-prey system,
constructed in the $\Delta^*$-$\delta$ parameter space, showing the
trapped (shaded) and escaped phases. The phase boundary (thick solid line)
is obtained analytically, and matches with the simulation data (boxes).
The horizontal thin dotted line ($\delta=1$) represents
the upper-bound for the trapped-to-escaped dynamical phase transition
(see text). (b) The dependence of the
catching range ($\Delta^*$) on the initial separation
(${r_0}^*$), as obtained from simulations.}
\end{figure}

In our simulations, the control parameter is the initial distance
$r_0$ between predator and prey. By tuning $\delta$ for a given $r_0$,
we actually found the
point of transition from trapped to escaped state.
The border-line case of steady hunting was found to be unstable.
Close to the border-line situation, the predator-prey distance $r_{12}$
remains constant for long times before fluctuations throw them into
either the trapped or the escaped state.
This constant distance, if identified with $\Delta$,
matches the phase boundary (Eq.(8)) perfectly, and $\Delta^*$
depends on $r_0^* (= r_0 / \Delta_0)$ in a roughly linear fashion (Fig.2(b)).

\begin{figure}
\includegraphics[width=10.0cm]{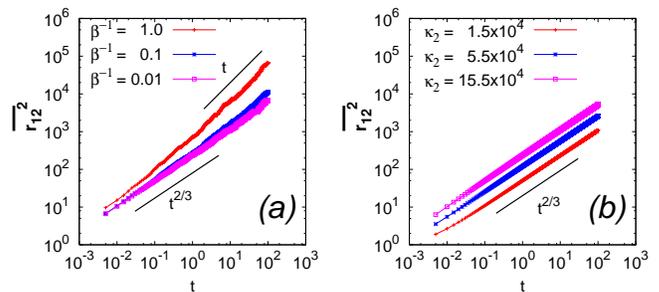}
\caption{\label{figMSD} (color online). Mean square displacement of the
prey w.r.t. the predator as a function of time, with
$\lambda_1 = 1$, $\gamma_1 = 10$, $\gamma_2 = 0.01$,
$D_{c1} = 10^3$, $D_{c2} = 10^2$, $\kappa_1 = 10^4$ 
for (a) $\beta^{-1} = 0.01,0.1,1.0$ and $\kappa_2 = 15.5 \times 10^4$;
and (b) zero noise case ($\beta^{-1} = 0$) and $\kappa_2 = 15.5 \times 10^4,
5.5 \times 10^4, 1.5 \times 10^4$. The power law behaviors $t^{2/3}$ and
$t$ are illustrated by the corresponding reference lines drawn.}
\end{figure}

What are the dynamical features of the escaped and the trapped phases?
Our simulations show that for escape (Fig.3(a)), the mean squared displacement
of the prey w.r.t. the predator gradually deviates from the initial
value $r_0$ and grows subdiffusively with time as
$\overline{ {{\bf r}_{12}}^{2}} \sim t^{\alpha}$, with an exponent $\alpha = 2/3$,
where the bar denotes an averaging over noise for a given $r_0$.
This behavior finally crosses over to diffusion,
$\overline{ {{\bf r}_{12}}^{2}} \sim t$,
for long times. The crossover time $t_{co}$
decreases with increasing fluctuation strength $\beta^{-1}$.
Zero-noise simulations show the
sub{\-}diffusive motion as the final long-time behavior, with the same exponent
$\alpha$ (Fig.3(b)). This means that within this phase, the hunting
process continues with a sub{\-}diffusive dynamics of the prey in the
comoving frame of the predator; but finally due to effects of fluctuation,
the prey diffuses away freely. 
It is therefore appropriate to look for a theoretical estimate of $\alpha$
within the noise-less ideal case. We note from Eq.(7) that at the advancing
predator position, the chemo-attractant profile behind the prey is of the form
$c_2 (|x|) = 
\lambda_2 /(4 \pi D_{c2} |x|) $ in the
steady state condition, since $x < 0$.
Solving the resulting equation of motion in steady state, 
$\gamma_1 {\dot x} = \kappa_1 |\nabla c_2 (x)| $, gives
$x^2 \sim t^{2/3}$ explaining the subdiffusive exponent.
When the fluctuations ultimately overcome the chemotactic coupling,
crossover to final diffusion results, requiring
$\beta^{-1} = \kappa_1 c_2 (|x|)
= \kappa_1 \lambda_2 / (4 \pi D_{c2} |x|)$. At the crossover point
from subdiffusive to the diffusive regime,
$x(t=t_{co}) \sim {t_{co}}^{1/3}$; implying that the crossover
time scales with the inverse effective temperature as $t_{co} \sim \beta^3 $.

At the front of each microbe,
the respective chemical profile decays much faster
$\sim \frac{1}{x} \exp{(-v_i x/D_{ci})}$.
The predator is thus always
at an advantage of sensing the prey from much longer distances.
Therefore, for very low $\delta$, a small increase in
$\delta$ greatly increases the catching range for the predator.
This accounts for the almost vanishing slope of the phase boundary for
low $\delta$ in the dynamical phase diagram (Fig.2(a)).
In this part of the phase diagram,
for two predators with close sensibility ratio, the one with the slightly
larger $\delta$ will successfully trap preys which started at a much
larger initial separation. For intermediate
values of $\delta$, the catching range $\Delta^*$ increases at a slower
rate with increase in $\delta$. This is because the initial separation
is large enough that small changes in $\Delta^*$ do not appreciably increase
the chemoattractant gradient ($\sim 1/x^2$) for the predator. One then needs
to considerably alter the sensibility ratio for obtaining a significant
change in the chemotactic coupling. 
As $\delta$ is further increased,
$\Delta^*$ increases to diverge at $\delta = 1$.

\begin{figure}
\includegraphics[width=10.0cm]{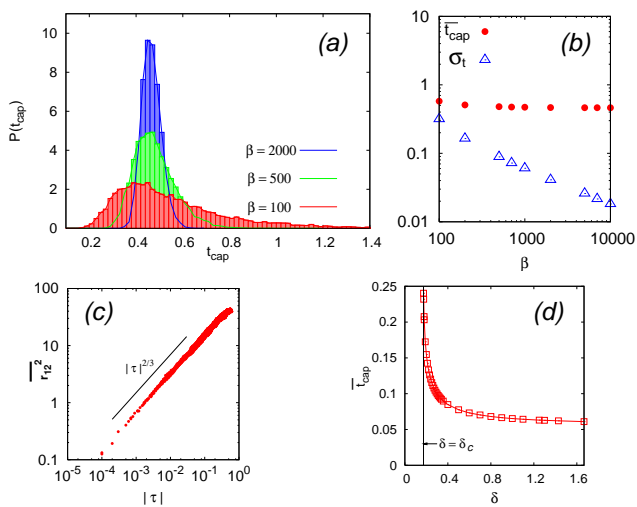}
\caption{\label{figChase} (color online).
(a) Probability distribution of the capture time for the
typical values of $\delta$ and $\Delta^*$ inside the trapped phase,
for $\beta = 100, 500, 2000$.
(b) Mean capture time ($\overline{t}_{cap}$) and variance ($\sigma_{t}$)
of the capture time, as a function of $\beta$.
(c) Mean square predator-prey separation
(${\overline{{{\bf r}_{12}}^2}}$), close to trapping situations, as a
function of the time interval $|\tau| = |t - t_{cap}|$.
The thick reference line
indicates $\tau^{2/3}$ power-law behavior.
(d) Divergence of the capture time with decrease in $\delta$, for a
fixed catching range ($\Delta^*$), as the trapped-to-escape transition
($\delta = \delta_c$, shown by the vertical line) is approached.}
\end{figure}

For the trapping situation we find a broad skewed distribution ($P(t_{cap})$)
of capture time $t_{cap}$, for a fixed $\beta$, $r_0$ and $\delta$ (Fig.4(a)).
The mean capture time $\overline{t}_{cap}$ is, however, independent of
the effective fluctuation strength, while the variance
$\sigma_t = [{\overline{(t_{cap} - \overline{t}_{cap})^2}}]^{1/2}$ decreases with
increasing $\beta$ (Fig.4(b)). The trapping
dynamics also show non-trivial power-law behavior, very close to capture:
${{\overline{{{\bf r}_{12}}^2}} (\tau \to 0-)} \sim |\tau|^{2/3}$, where $\tau = t-t_{cap}$
(Fig.4(c)).
We note that initially, close to $t=0$, when the predator-prey distance is
$\sim r_0$, fluctuations dominantly control the individual microbe's
motion until a steady chemical concentration profile sets up in the
process to favour a systematic dynamics. This time scale is dependent
on the individual diffusivities of the chemical: higher the chemical
diffusivities compared to the non-chemotactic diffusivities of the microbes,
the faster will the predator-prey chemotactic systematics set in. For the
trapping dynamics very close to capture, therefore, the predator is already
responding to the steady chemoattractant gradient
$\gamma_1 {\dot x} = \kappa_1 |\nabla c_2(r) |_{r=x} \sim 1/{x}^2$.
Thus, integrating from $x=x(t)$ to $x = x(t=t_{cap}) = 0$, we obtain
the scaling form $x^2 \sim (t_{cap}-t)^{2/3}$, with $t < t_{cap}$.
For fixed initial separation, $t_{cap}$ diverges on approaching from above
the corresponding critical value of the sensibility ratio ($\delta_c$)
for escape (Fig.4(d)).

In conclusion, we studied the dynamics of simple gradient sensing
chemotactic microorganisms in a predator-prey relationship.
Although prokaryotes like most bacteria chemotax chiefly through
temporal comparison
of chemical gradients, the direct spatial sensing which we studied here
is prevalent among eukaryotes like amoeba, yeast cells,
neutrophils, lymphocytes and glial cells.
Here we have been able to delineate power law behaviors in the chasing process
both from theory and numerical simulations, and crossover time scaling
with fluctuation strength.
A dynamical phase diagram has been obtained
to identify conditions for escape and catching, with a border-line
unstable steady-hunting situation. A broad class of microbes varying widely
with respect to their mobility, secretion rates and diffusivities of
ejected chemicals and strength of spatial gradient sensing of chemicals
can be located in the phase diagram. Interestingly, a simple sensibility
ratio resulting from our model calculation allows a simple criterion to predict
the outcome of such hunting processes: a trapped or an escaped situation
depending on the initial predator-prey separation, and a no escape situation
independent of their initial separation.


We thank S. van Teeffelen, P. Romanczuk, P.S. Hammond and R. Winkler
for helpful discussions.
This work was supported by DFG within SFB TR6 (project D3).

\bibliographystyle{apsrev}

\end{document}